\documentclass[conference]{IEEEtran}
\IEEEoverridecommandlockouts
\usepackage{cite}
\usepackage{amsmath,amssymb,amsfonts}
\usepackage{algorithmic}
\usepackage{graphicx}
\usepackage{textcomp}
\usepackage{xcolor}
\usepackage{booktabs}
\usepackage{placeins}

\renewcommand{\textbf}[1]{#1}
\renewcommand{\emph}[1]{#1}

\def\BibTeX{{\rm B\kern-.05em{\sc i\kern-.025em b}\kern-.08em
    T\kern-.1667em\lower.7ex\hbox{E}\kern-.125emX}}
\begin{document}

\title{QCL-IDS: Quantum Continual Learning for Intrusion Detection with Fidelity-Anchored Stability and Generative Replay\\
}

\author{\IEEEauthorblockN{Zirui Zhu}
\IEEEauthorblockA{\textit{Johns Hopkins University} \\
\textit{Information Security Institute}\\
Baltimore, USA \\
ziruizhu87@gmail.com}
\and
\IEEEauthorblockN{Xiangyang Li}
\IEEEauthorblockA{\textit{Johns Hopkins University} \\
\textit{Information Security Institute}\\
Baltimore, USA \\
xyli@jhu.edu}
}

\maketitle

\begin{abstract}
Continual intrusion detection must absorb newly emerging attack stages while retaining legacy detection capability under strict operational constraints: bounded compute/qubit budgets and privacy rules that preclude long-term storage of raw telemetry. We propose \emph{QCL-IDS}, a quantum-centric continual-learning framework that co-designs \emph{stability} and \emph{privacy-governed rehearsal} for NISQ-era pipelines. Its core component, \emph{Q-FISH (Quantum Fisher Anchors)}, enforces retention using a compact anchor coreset by (i) sensitivity-weighted parameter constraints and (ii) a fidelity-based functional anchoring term that directly limits decision drift on representative historical traffic. To regain plasticity without retaining sensitive flows, QCL-IDS further introduces privacy-preserved quantum generative replay (QGR) via frozen, task-conditioned generator snapshots that synthesize bounded rehearsal samples. Across a three-stage attack stream on \textbf{UNSW-NB15} and \textbf{CICIDS2017}, QCL-IDS consistently attains the best retention--adaptation trade-off: the gradient-anchor configuration achieves \textbf{mean Attack-F1 = 0.941 with forgetting = 0.005} on UNSW-NB15 and \textbf{mean Attack-F1 = 0.944 with forgetting = 0.004} on CICIDS2017, versus \textbf{0.800/0.138} and \textbf{0.803/0.128} for sequential fine-tuning, respectively.
\end{abstract}


\section{Introduction}
Intrusion detection systems (IDS) are not static classifiers but continually maintained capabilities. As enterprise networks evolve, IDS must assimilate new attack stages without erasing the decision boundaries required to detect historical threats \cite{garcia2009anomaly,sommer2010closedworld,buczak2016survey,gama2014conceptdrift}. This challenge is compounded by operational realities: updates must be deployed under bounded compute budgets, and strict data privacy governance often forbids the long-term retention of raw network telemetry required by traditional replay buffers \cite{rebuffi2017icarl,lopezpaz2017gem,chaudhry2019tiny,dwork2014algorithmic}.

Existing classical continual learning methods often fail to reconcile these constraints. Replay-based methods compromise privacy by storing raw samples, while regularization methods often rely on parameter-space approximations that struggle to capture the true functional sensitivity of the model \cite{rebuffi2017icarl,lopezpaz2017gem,kirkpatrick2017ewc,zenke2017synaptic,aljundi2018mas,huszar2018note}.

\subsection*{Why Quantum}
We posit that Variational Quantum Circuits (VQCs) offer a distinct structural advantage for this constrained setting: efficient geometric regularization. In classical deep networks, accurately calculating the Fisher Information Matrix to diagnose parameter importance is computationally prohibitive ($O(N^2)$). In contrast, the quantum state space allows us to compute Quantum Fisher Information (QFI) and Fidelity—measures of geometric distance in the Hilbert space—efficiently via parameter-shift rules \cite{amari1998natural,martens2015kfac,braunstein1994statistical,jozsa1994fidelity,schuld2019evaluating,stokes2020qng,preskill2018nisq}. This allows us to regularize the model based on how the quantum state changes, providing a far more rigorous stability guarantee than Euclidean weight penalties.

We use a QFI-inspired sensitivity surrogate and a fidelity-inspired functional anchoring term, both of which can be estimated with only a small number of circuit evaluations (via parameter-shift and/or finite-difference approximations), making the approach well aligned with the tight evaluation budgets of NISQ hardware.

\subsection*{Contributions}
We introduce QCL-IDS, a framework co-designed for operational feasibility and privacy compliance. Our specific contributions are:
\begin{itemize}
\item \textbf{Q-FISH with Fidelity constraints.} Moving beyond standard EWC, we introduce Q-FISH, which anchors the model's evolution using Quantum Fisher Information estimated on small, representative coresets. Crucially, we incorporate a Fidelity penalty that explicitly constrains the quantum state overlap between updates. This ensures that the model preserves its decision geometry on critical samples  rather than just preserving weight values.

\item \textbf{Privacy-Preserved QGR.} To solve the data-retention problem, we introduce a budgeted replay mechanism using frozen generator snapshots. Instead of storing privacy-sensitive traffic logs, we train compact quantum generators to approximate the distribution of past tasks. This allows the system to rehearse historical knowledge without violating data governance policies regarding raw telemetry retention.

\item \textbf{Budgeted Capability Expansion.} We formulate continual IDS as a resource-constrained optimization problem. Evaluations on UNSW-NB15 confirm that stability regularization via Q-FISH is the primary driver of retention, while our privacy-aligned replay provides the necessary plasticity to learn new attack vectors without catastrophic forgetting.

\end{itemize}

\section{Methodology}
\label{sec:method}

This section describes \emph{QCL-IDS}, a quantum-centric continual learning framework for intrusion detection under a stream of evolving tasks. While the discriminative and generative cores are implemented as quantum circuits, QCL-IDS follows the standard NISQ training loop with classical optimization and classical preprocessing; these classical components serve as the execution scaffold and controlled baselines rather than the primary methodological contribution \cite{preskill2018nisq,biamonte2017quantum,benedetti2019parameterized}.
The design targets a practical constraint common in security pipelines: models must be updated online as new attack stages appear, while preserving detection capability for previously observed stages without storing the full historical dataset.

\paragraph{High-level idea}
At each task $t$, QCL-IDS trains a compact quantum classifier on a mixture of (i) the current-task data and (ii) replayed samples synthesized from previously learned tasks. To prevent destructive drift, the classifier is stabilized by an anchor-based regularizer (Q-FISH) \cite{shin2017dgr,rebuffi2017icarl,chaudhry2019tiny} computed on a small coreset. The framework is modular and comprises three components:
(i) \textbf{Q-FISH} anchor-based stability regularization,
(ii) \textbf{Quantum generative replay (QGR)} using task-conditioned generator snapshots (with a classical generator as a control), and
(iii) a \textbf{quantum classifier backbone} implemented as a data re-uploading variational quantum circuit (VQC).

\begin{figure}[!ht]
    \centering
    \includegraphics[width=1\linewidth]{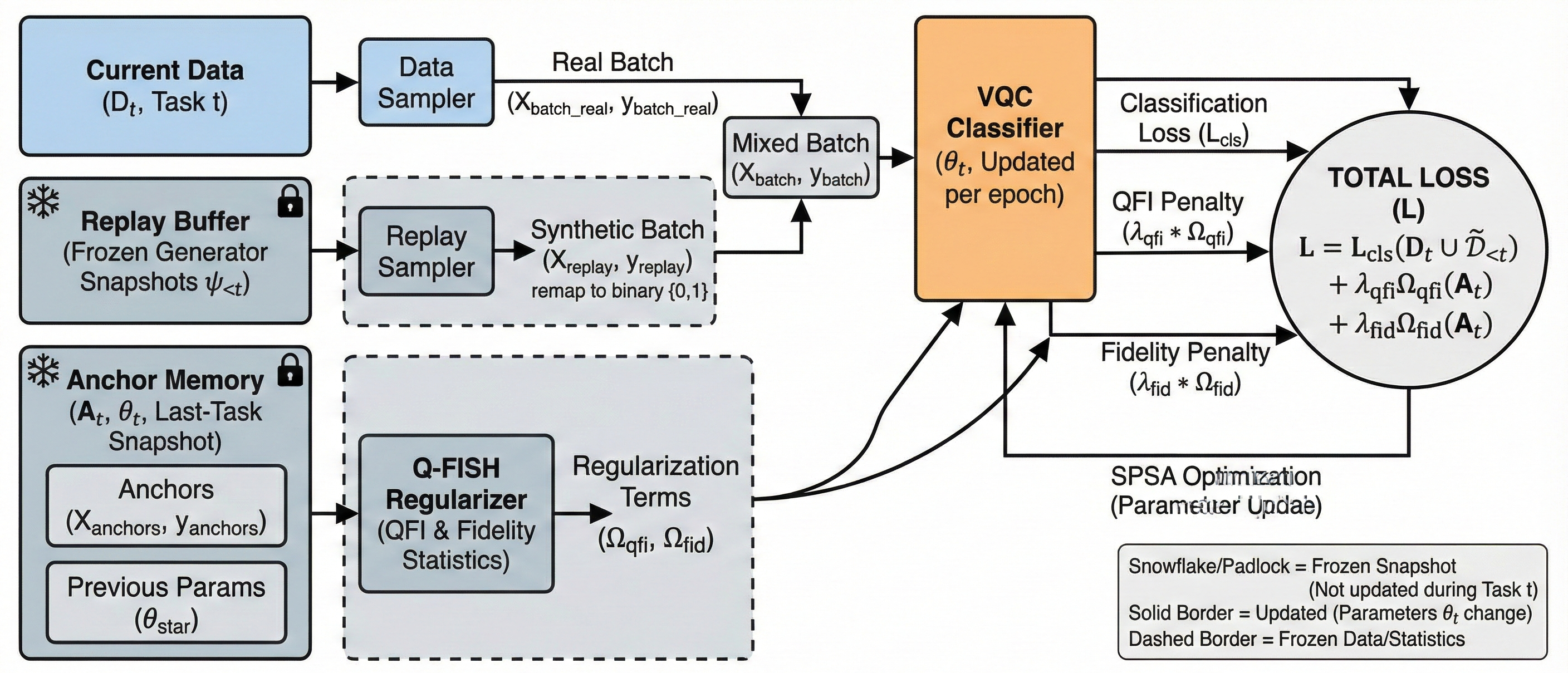}
    \caption{System overview}
    \label{fig:system_overview}
\end{figure}

\subsection{Continual Learning Setup and Persistent State}
\label{sec:cl_setup}

\paragraph{Task stream}
We consider a sequence of $T$ tasks $\{\mathcal{T}_t\}_{t=0}^{T}$. Each task provides labeled data
$\mathcal{D}_t=\{(x_i,y_i)\}_{i=1}^{n_t}$, where $y\in\{0,1\}$ denotes benign vs. attack for the current task definition.
Each input $x\in\mathbb{R}^d$ is a compact continuous vector produced by the upstream preprocessing/encoding stage (details in the experimental setup). Before quantum encoding, a lightweight conditioning layer maps $x$ to a clipped angle vector $\tilde{x}\in[-1,1]^q$ (linearly rescaled to rotation angles), where $q$ equals the number of qubits in the VQC and thus the number of encoded coordinates per re-uploading layer.

\paragraph{Goal}
After learning task $\mathcal{T}_t$, the classifier should (i) achieve strong predictive performance on $\mathcal{D}_t$ and (ii) retain performance on all prior tasks $\{\mathcal{T}_k\}_{k<t}$, mitigating catastrophic forgetting.

\paragraph{State carried across tasks}
QCL-IDS maintains three persistent states:
\begin{itemize}
  \item \textbf{Classifier state:} parameters $\theta_t$ of the quantum classifier backbone (Section~\ref{sec:q_backbone}).
  \item \textbf{Replay state:} a set of frozen generator snapshots $\{\psi_k\}_{k<t}$ that synthesize replay data for past tasks (Section~\ref{sec:qgr}).
  \item \textbf{Regularization state:} compact anchor memory $\{\mathcal{A}_k\}_{k<t}$ with the quantities needed by Q-FISH (e.g., parameter snapshots and anchor statistics; Section~\ref{sec:qfish}).
\end{itemize}
This structured memory replaces storing full historical datasets and keeps the continual footprint bounded.

\subsection{Per-task Training Objective and Workflow}
\label{sec:cl_objective_workflow}

\paragraph{Composite objective}
At task $t$, QCL-IDS optimizes a single objective that combines current-task supervision, replay rehearsal, and stability regularization:
\begin{equation}
\label{eq:cl_objective_core}
\mathcal{L}_t(\theta)=
\mathcal{L}^{\text{sup}}_t(\theta)
\;+\;
\alpha\,\mathcal{L}^{\text{replay}}_t(\theta)
\;+\;
\mathcal{R}_{<t}(\theta),
\end{equation}
where $\mathcal{L}^{\text{sup}}_t$ is the standard supervised loss on $\mathcal{D}_t$, $\mathcal{L}^{\text{replay}}_t$ is the same loss evaluated on synthesized replay samples from prior tasks, and $\mathcal{R}_{<t}$ is the Q-FISH stability regularizer computed using anchors from past tasks. The scalars $\alpha$ and $\lambda$ control the replay and stability strengths, respectively.

\paragraph{Continual workflow}
For each task $\mathcal{T}_t$, QCL-IDS executes the following steps:
\begin{enumerate}
  \item \textbf{Replay construction:} sample a replay set from the frozen generator snapshots $\{\psi_k\}_{k<t}$ and mix it with $\mathcal{D}_t$ according to the replay policy (Section~\ref{sec:qgr_policy}).
  \item \textbf{Classifier update:} update $\theta$ by minimizing Eq.~\eqref{eq:cl_objective_core}, where Q-FISH constrains drift on past anchors (Section~\ref{sec:qfish}).
  \item \textbf{Memory update:} after training, (i) train a new generator snapshot $\psi_t$ for task $t$ and freeze it for future replay, and (ii) construct a compact anchor set $\mathcal{A}_t$ and store the statistics required by Q-FISH.
\end{enumerate}

\begin{figure}[!ht]
    \centering
    \includegraphics[width=1.0\linewidth]{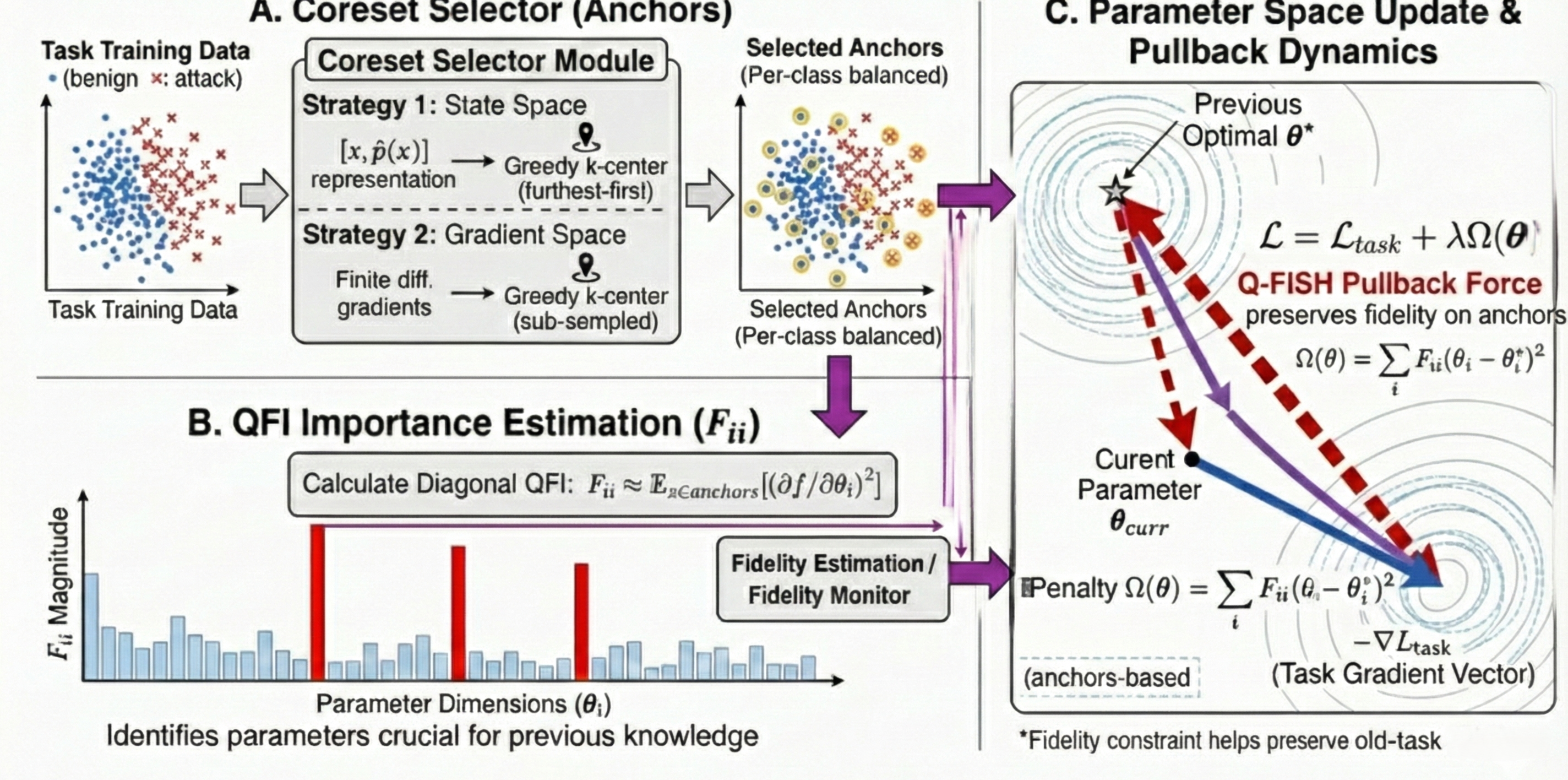}
    \caption{Q-FISH concept schematic}
    \label{fig:qfish_concept}
\end{figure}
\subsection{Q-FISH: Anchor-based Stability Regularization}
\label{sec:qfish}

To mitigate catastrophic forgetting, we introduce Q-FISH, a regularization mechanism that constrains the model's functional drift on a set of representative anchor samples from prior tasks. Unlike approaches that rely on weight-space proximity alone, Q-FISH explicitly penalizes deviations in the model's scalar output (expectation values) using a composite objective.

Let $\theta^*$ denote the parameters frozen after the previous task. The regularization term $\mathcal{R}_{<t}(\theta)$ is defined as:

\begin{equation}
\label{eq:qfish_total}
\begin{aligned}
\mathcal{R}_{<t}(\theta)
&= \lambda_{qfi} \sum_{i} \hat{F}_{ii}\,(\theta_i - \theta_i^*)^2 \\
&\quad + \lambda_{fid}\!\left(
1 - \frac{1}{|\mathcal{A}_{fid}|} \sum_{x \in \mathcal{A}_{fid}} \widehat{fid}(x)
\right).
\end{aligned}
\end{equation}

This objective comprises two complementary terms computed on batches $\mathcal{A}_{qfi}$ and $\mathcal{A}_{fid}$, which are sampled from the historical coreset $\mathcal{A}_{<t}$:

\subsubsection*{Sensitivity-Weighted Parameter Constraints}
The first term is a quadratic penalty on parameter changes, weighted by a diagonal sensitivity proxy $\hat{F}_{ii}$. Rather than computing the full Quantum Fisher Information Matrix or the Classical Fisher Information (which requires computing gradients over the output distribution), we estimate the importance of each parameter $\theta_i$ by measuring the sensitivity of the model's scalar output expectation $f_\theta(x)$ via finite differences:

\begin{equation}
\hat{F}_{ii} \approx \frac{1}{|\mathcal{A}_{qfi}|} \sum_{x \in \mathcal{A}_{qfi}} \left( \frac{f_{\theta+\epsilon e_i}(x) - f_{\theta-\epsilon e_i}(x)}{2\epsilon} \right)^2
\label{eq:qfi_ewc_core}
\end{equation}

where $\epsilon$ is a small perturbation hyperparameter and $e_i$ is the unit vector for the $i$-th parameter. Parameters that cause significant variations in the model's predictions on historical anchors yield high $\hat{F}_{ii}$ values and are thus heavily regularized in Eq. \ref{eq:qfish_total}.

\subsubsection*{Functional Fidelity Proxy}
The second term explicitly discourages drift in the decision space. We define a bounded fidelity proxy $\widehat{fid}(x)$ based on the squared difference between the current model output $f_\theta(x)$ and the frozen snapshot output $f_{\theta^*}(x)$:

\begin{equation}
\widehat{fid}(x) = \text{clip}\left( 1 - \frac{1}{2}(f_\theta(x) - f_{\theta^*}(x))^2, 0, 1 \right)
\label{eq:fid_proxy}
\end{equation}

This term acts as a direct functional anchor, ensuring that the classifier's predictive behavior on critical past samples remains consistent with the previously learned state $\theta^*$. Note that $\mathcal{A}_{qfi}$ and $\mathcal{A}_{fid}$ may be distinct sub-samples from the coreset $\mathcal{A}_{<t}$ to optimize computational efficiency during the update loop.

\begin{figure}[!ht]
    \centering
    \includegraphics[width=1\linewidth]{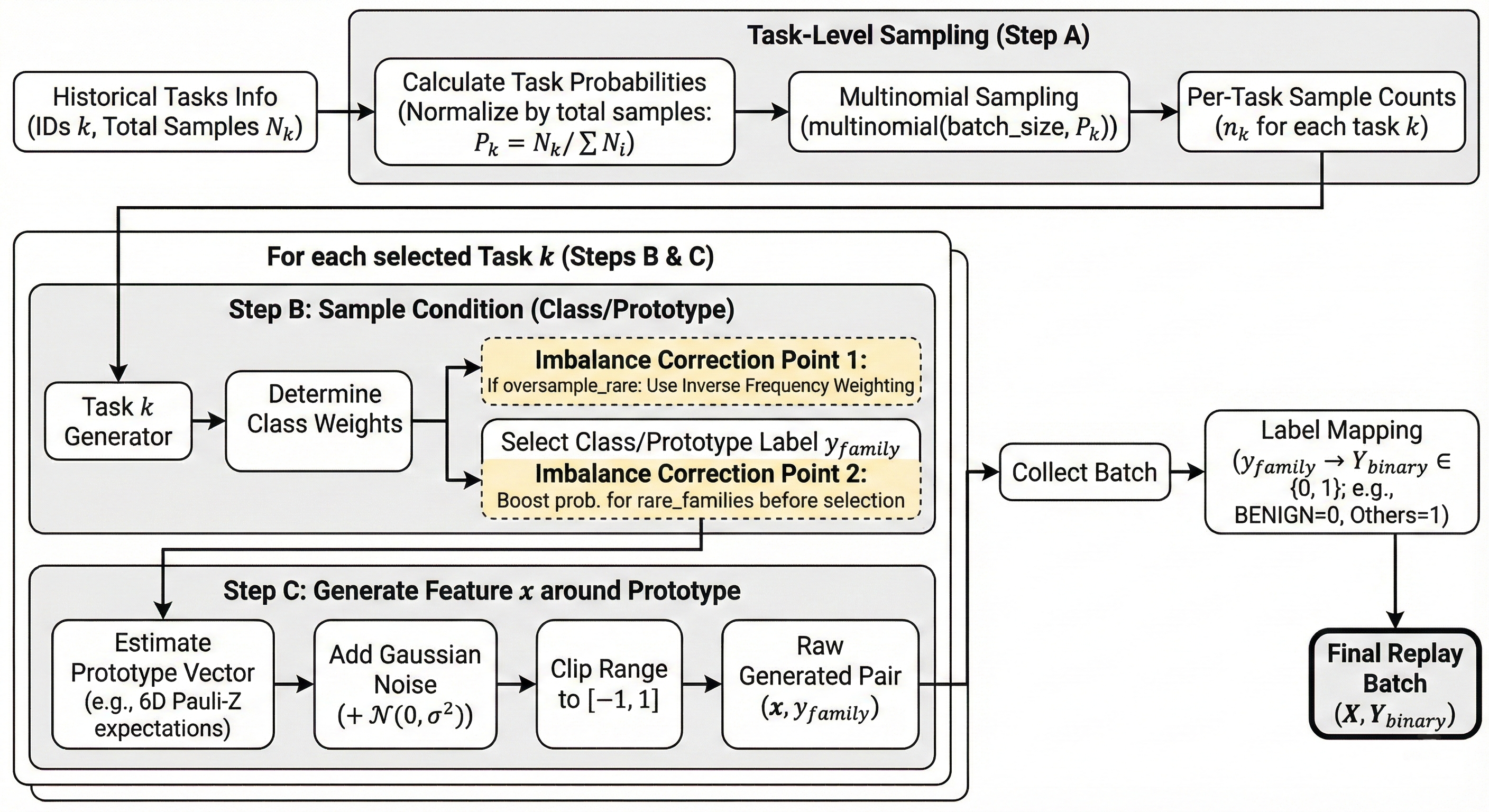}
    \caption{Replay sampling policy diagram}
    \label{fig:replay_policy}
\end{figure}
\subsection{Quantum Generative Replay}
\label{sec:qgr}

Generative replay mitigates forgetting by approximating past task data distributions and synthesizing rehearsal samples during later tasks. QCL-IDS maintains a set of \emph{frozen} generator snapshots, one per task, to avoid interference between the generative models themselves.

\subsubsection{Quantum Prototype Learning}
For each task, we train a parameterized quantum generator $\mathcal{G}_{\phi}$ to encode the statistical centroids of the data classes. Let $\mu_c$ denote the empirical mean vector (prototype) for class $c$ computed from the current task data. The generator parameters $\phi$ are optimized using the SPSA optimizer \cite{spall1992multivariate} to minimize the Euclidean distance between the circuit's measured expectation values (observable outputs) and the target prototypes:

\begin{equation}
\mathcal{L}_{gen}(\phi) = \sum_{c} \left\| f_{\phi}(c) - \mu_c \right\|^2
\label{eq:gen_loss}
\end{equation}

where $f_{\phi}(c)$ represents the expectation value of the quantum circuit observables given condition $c$. This formulation allows the high-dimensional quantum state space to efficiently compress and store the structural centers of the historical data manifolds.

\subsubsection{Hybrid Replay Synthesis}

During the training of subsequent tasks, we reconstruct historical data by combining the stored quantum information with classical stochastic diffusion. A synthetic replay sample $x_{replay}$ is generated in a two-step process:

\begin{equation}
x_{replay} = f_{\phi^*}(c) + \xi, \quad \text{where } \xi \sim \mathcal{N}(0, \sigma^2 I)
\label{eq:replay_synthesis}
\end{equation}

First, the frozen quantum generator $\phi^*$ is queried to recover the prototype $f_{\phi^*}(c)$. Second, isotropic Gaussian noise $\xi$ is added classically to model the intra-class variance. This hybrid factorization ensures computational efficiency, as the quantum circuit is dedicated solely to maintaining the decision-critical centroids, while the distribution spread is handled via classical post-processing.

\subsubsection{Replay policy}
\label{sec:qgr_policy}
During task t, replay samples
are drawn by first selecting a past task snapshot, then selecting
a condition (and mixture component, if applicable), and finally
sampling synthetic pairs (x,y) for rehearsal. The policy can be
configured to: (i) balance tasks to avoid over-fitting the most
recent task, (ii) mitigate class imbalance, and (iii) upweight
rare but security-critical modes. We treat the sampling policy
as part of the experimental protocol to ensure reproducibility
and fair comparisons.
\subsection{Quantum Classifier Backbone}
\label{sec:q_backbone}

The discriminative backbone is a data re-uploading VQC that maps compact continuous features into a quantum feature space and outputs a calibrated attack probability. The circuit definition is decoupled from the continual-learning logic: replay affects the training data distribution, and Q-FISH affects the regularization term, without altering the circuit structure \cite{perezsalinas2020data,mitarai2018quantum,benedetti2019parameterized}.
\begin{figure}[!ht]
    \centering
    \includegraphics[width=1\linewidth]{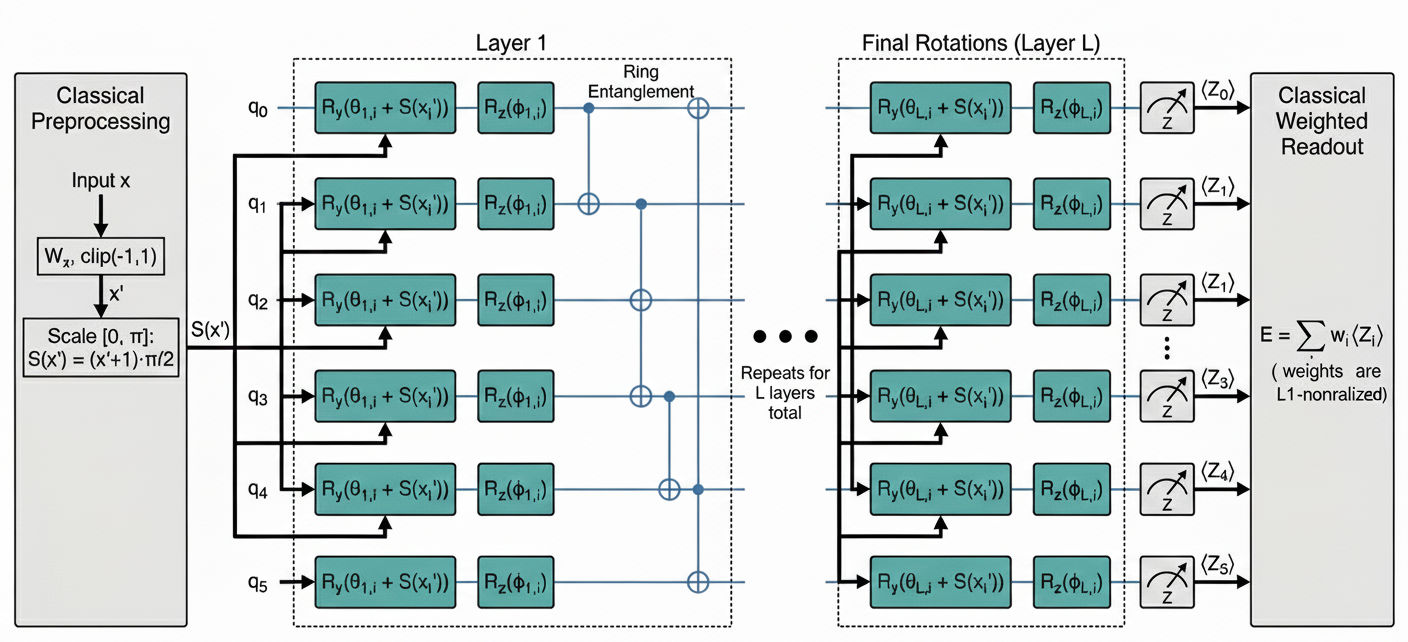}
    \caption{Data re-uploading VQC circuit}
    \label{fig:vqc_circuit}
\end{figure}
\subsubsection{Input conditioning and data re-uploading}
\label{sec:q_backbone_encoding}

Given an input vector $x$, the backbone (i) applies a lightweight trainable conditioning layer (e.g., a linear projection with clipping) to align the input with the valid quantum encoding range, and (ii) injects the resulting angles repeatedly through multiple re-uploading layers interleaved with trainable rotations and entangling gates. This re-uploading pattern increases expressivity under a fixed qubit budget by allowing the circuit to represent higher-order feature interactions using a shallow, repeating structure. Exact operator forms are provided in Appendix~\ref{app:math_details}.

\subsubsection{Readout and probability calibration}
\label{sec:q_backbone_readout}

The circuit produces a scalar score using a learnable linear readout over Pauli observables (typically $Z$ expectations), and maps this score to a probability via a calibrated sigmoid:
\begin{equation}
\label{eq:readout_prob_core}
p_{\theta}(y{=}1\mid x)=\sigma\!\Big(\tau\big(\sum_{i=1}^{q} w_i\,\langle Z_i\rangle - b\big)\Big),
\end{equation}
where $w_i$ are learnable readout weights, $b$ is a bias term, and $\tau$ is a temperature-like calibration parameter. 
Unless otherwise stated, we optimize the calibration temperature $\tau$ jointly with $(w,b,\theta)$ as a constrained positive scalar, effectively treating Eq.~\eqref{eq:readout_prob_core} as an explicit one-parameter calibration layer integrated into training \cite{guo2017calibration}.
This explicit calibration is useful in security settings where threshold selection and operating points (e.g., low false-positive regimes) are operationally important.

\subsubsection{Optimization interface}
\label{sec:q_backbone_opt}

QCL-IDS supports optimizers appropriate for quantum evaluation, including gradient-free methods (e.g., SPSA) and gradient-based updates when analytic gradients are available. Importantly, the continual pipeline remains unchanged across optimizers: Eq.~\eqref{eq:cl_objective_core} defines the training objective, while QGR and Q-FISH provide replay data and stability signals through well-defined module interfaces \cite{schuld2019evaluating}.

\subsection{Summary of Module Interfaces}
\label{sec:interfaces_summary}

For clarity and reproducibility, we summarize the inputs/outputs of each module:
\begin{itemize}
  \item \textbf{Classifier (VQC)} takes $(x,y)$ mini-batches and returns predicted probabilities; it exposes parameters $\theta$ for optimization.
  \item \textbf{Replay generator} stores task snapshots $\{\psi_t\}$ and provides a sampler that emits synthetic labeled pairs $(x,y)$ under a specified policy.
  \item \textbf{Q-FISH} stores anchors $\{\mathcal{A}_t\}$, parameter snapshots $\{\theta_t\}$, and anchor statistics; it returns the stability penalty $\mathcal{R}_{<t}(\theta)$ during training.
\end{itemize}
\section{Experimental Setup}
\label{sec:exp_setup}

\subsection{Datasets and Attack-Stage Continual Learning Tasks}
\label{sec:exp_datasets_tasks}

We evaluate on two intrusion detection benchmarks, CICIDS2017 and UNSW-NB15 \cite{sharafaldin2018cicids,moustafa2015unsw}, under an \emph{attack-stage} continual learning protocol.
Each dataset is mapped into the same sequence of three binary tasks:
(i) \texttt{NORMAL} vs \texttt{RECON\_SCAN},
(ii) \texttt{NORMAL} vs \texttt{DOS\_RESOURCE},
and (iii) \texttt{NORMAL} vs \texttt{INTRUSION\_MALWARE}.
The phase mapping is implemented as code-defined label-to-phase maps (one per dataset); unknown labels trigger an explicit error to keep the task definition verifiable and reproducible.

\paragraph{Train/validation/test splitting and leakage control}
For each phase-specific task, attack flows are randomly shuffled and partitioned \emph{by count} into train/validation/test using a $0.6/0.2/0.2$ ratio (no stratification is applied; within a task, the attack side contains a single phase and thus phase-stratification is not applicable).
Splits are constructed \emph{independently per task}: each task first partitions its attack subset, then draws NORMAL samples as needed for that task.
NORMAL flows are drawn from a single globally shuffled pool (random\_state=42) using a disjoint sequential allocation scheme.
Concretely, we maintain a running pointer \texttt{normal\_used} into the shuffled NORMAL pool; for task $k$, we compute the required NORMAL counts for train/validation/test to match the dataset-level normal:attack ratio, take a contiguous slice from the pool, split it into train/validation/test by the computed counts, and advance \texttt{normal\_used}.
This guarantees that NORMAL samples are disjoint across tasks and disjoint across splits.
Attack samples are naturally disjoint across tasks since tasks correspond to non-overlapping phase labels (\texttt{RECON\_SCAN}, \texttt{DOS\_RESOURCE}, \texttt{INTRUSION\_MALWARE}).
No explicit deduplication or near-duplicate removal is applied during splitting; only missing/\texttt{inf} handling and row filtering are performed.

\subsection{Feature Budget and Preprocessing}
\label{sec:exp_preprocessing}

All inputs to the quantum models are represented as \textbf{6-dimensional} feature vectors and normalized into the interval $[-1,1]$ prior to quantum encoding.
Unless stated otherwise, both datasets use \emph{numeric features $\rightarrow$ PCA(6) $\rightarrow$ 6D} to match the fixed quantum input budget.

\paragraph{Numeric-only feature selection}
We retain numeric columns only; non-numeric/categorical fields are discarded (no one-hot encoding or embeddings).
During loading, $\pm\infty$ values are replaced with NaN and rows containing NaNs in feature columns are removed (UNSW-NB15 additionally drops rows with missing \texttt{attack\_cat} prior to feature extraction).
As a safety measure, the standardization pipeline includes median imputation, although in practice most invalid rows are filtered earlier.

\paragraph{Per-task standardization and scaling}
For each task, we fit a \texttt{StandardScaler} on the task's training split only and apply the same transformation to validation/test.
We then linearly rescale each standardized feature dimension to $[-1,1]$ using the \emph{training-split} min/max for that task.
For near-constant dimensions with $(\max-\min) < 10^{-6}$ on the training split, the rescaled value is set to $0$ to avoid numerical instability.

\paragraph{Per-task PCA}
Dimensionality reduction is performed \emph{per task} by fitting \texttt{sklearn.decomposition.PCA(n\_components=6, random\_state=42)} on the current task's standardized training features (\texttt{whiten=False}, \texttt{svd\_solver='auto'} by default), then transforming the corresponding validation/test splits.
PCA is not shared across tasks and is not updated incrementally.

\subsection{Compared Methods and Ablation Groups}
\label{sec:exp_baselines}

We report results for the following experimental groups, as organized by the main experiment scripts:
\begin{itemize}
  \item \textbf{Baseline:} sequential training without replay and without Q-FISH/EWC regularization.
  \item \textbf{Conditional GMM Replay Only:} classical conditional GMM replay as a generator control.
  \item \textbf{Replay Only:} enable generative replay while disabling Q-FISH/EWC.
  \item \textbf{EWC Only:} enable the QFI drift term only while disabling replay.
  \item \textbf{Q-FISH Only:} enable Q-FISH regularization while disabling replay.
  \item \textbf{Full (state-space anchors):} replay + Q-FISH using \texttt{state\_space} coreset anchors.
  \item \textbf{Full (gradient-space anchors):} replay + Q-FISH using \texttt{gradient\_space} coreset anchors.
\end{itemize}
Some groups may be toggled on/off in scripts for runtime reasons, but the method and logging support all configurations.

\paragraph{Matched-feature comparisons}
Within a given script run, all experimental groups (including the logistic-regression oracle baseline used for intransigence) reuse the same cached \texttt{TaskSplit} and the same 6D representation.
By default, CICIDS2017 and UNSW-NB15 use numeric features followed by per-task PCA(6).
For CICIDS2017 only, an optional flag (\texttt{--no-pca-features}) switches to a fixed 6-feature hand-crafted subset; in that case, the same per-task standardization and $[-1,1]$ rescaling are still applied.
Unless stated otherwise, all reported comparisons are matched-feature and matched-split.

\subsection{Training Protocol and Default Hyperparameters}
\label{sec:exp_training}

\paragraph{Per-task training loop}
Each task is trained sequentially for a fixed number of epochs with mini-batches that may mix real current-task samples and replay samples (when replay is enabled).
Unless stated otherwise, the default batch size is $256$ and the default number of epochs per task is $20$.
No early stopping is used; all tasks run for the configured fixed epoch budget.

\paragraph{Quantum classifier configuration}
The VQC backbone uses $6$ input features aligned to $6$ qubits and a fixed circuit depth (default: \texttt{n\_layers=3} in the experiment scripts).
Shot-based evaluation is used (default: $2048$ shots).

\paragraph{Quantum generative replay configuration}
The default replay generator is the prototype-based ConditionalQCBM (6 qubits) with a shallow circuit depth (default: $2$ layers) and shot-based sampling (default: $1024$ shots).
The generator is trained with a fixed maximum number of optimization iterations (default: $300$ in scripts).

\paragraph{Replay--regularization coupling}
When replay is used alone (or regularization is used alone), the default replay ratio is $0.3$.
When \emph{both} replay and Q-FISH are enabled, the scripts reduce replay intensity and regularization weights to avoid over-regularization, using:
\[
\begin{cases}
\texttt{replay\_ratio: } 0.3 \rightarrow 0.1 \\
\lambda_{\mathrm{qfi}}: 0.3 \rightarrow 0.25 \\
\lambda_{\mathrm{fid}}: 0.1 \rightarrow 0.08
\end{cases}
\]

\paragraph{Class-imbalance handling}
Class imbalance is handled via weighted BCE or focal loss (configurable), with dataset-specific default strategies:
CICIDS2017 uses a square-root weighting strategy with an additional attack-class weight boost, while UNSW-NB15 uses an automatic strategy that selects recommended weights and/or focal loss based on detected imbalance.

\subsection{Noise Simulation and Hardware-Friendly Options}
\label{sec:exp_noise}

By default, experiments run under shot-based Aer noise simulation to approximate NISQ execution. Unless stated otherwise, the noise model is enabled for both training and evaluation (\texttt{--aer-noise}), the default parameters are \texttt{noise\_1q=0.001}, \texttt{noise\_2q=0.01}, \texttt{readout\_error=0.02}, and \texttt{aer\_method=\{density\_matrix\}} \cite{preskill2018nisq,abraham2019qiskit}.
The noise model is applied consistently to both training and evaluation, affecting VQC estimation and QCBM sampling.

\subsection{Evaluation Protocol and Metrics}
\label{sec:exp_metrics}

\paragraph{Classification metrics}
Because NORMAL often dominates, the primary metric is \textbf{Attack-F1} (F1 for class $y{=}1$).
We additionally report Accuracy, F1-macro, F1-weighted, Attack-Precision, Attack-Recall, and ROC-AUC when available.

\paragraph{Continual learning metrics}
We maintain a task-performance matrix $R[t][k]$ (performance on task $k$ after training through task $t$) and compute standard CL summaries including:
mean forgetting, mean backward transfer (BWT), forward transfer, and intransigence relative to a lightweight oracle baseline (logistic regression).

\paragraph{Per-phase analysis}
Because each example carries a phase label, we also report per-phase recall and family-level variants of the $R$ matrix and forgetting/BWT analyses.

\paragraph{Decision threshold selection}
Each task uses a task-specific threshold $\tau_k$ selected by maximizing F1-macro on a held-out subset of the training data.
Concretely, for each task we randomly sample a threshold-tuning subset of size $\max(100,\;0.2|X_{\mathrm{train}}|)$ from the training split without stratification, and select $\tau_k$ via a post-hoc search.
Evaluation uses each task's own $\tau_k$ rather than a fixed $0.5$ threshold.
A \texttt{class\_incremental} option controls whether thresholds for past tasks may be re-optimized, enabling either (i) stricter accounting of forgetting or (ii) a more deployment-aligned adaptive-threshold protocol.
No additional probability calibration is performed beyond the model's native output mapping.

\subsection{Implementation and Reproducibility}
\label{sec:exp_repro}

All experiments are executed via two entry-point scripts (one per dataset), which log metrics and figures to dataset-specific result folders.
Task splits are cached in a versioned processed-data directory to ensure deterministic reuse of the same task stream across runs.

\paragraph{Run protocol and stochasticity}
The current implementation reports single-run results (no multi-seed aggregation).
Dataset splitting uses fixed seeds (base seed $42$ with task-index offsets), and PCA uses \texttt{random\_state=42}.
However, training-time minibatch formation, replay sampling, and threshold-subset sampling rely on \texttt{numpy} random draws that are not globally seeded in the default scripts; therefore results are reproducible at the split level but not guaranteed to be bit-level deterministic unless a global NumPy seed is set.

\paragraph{Software versions}
Python 3.12.12; Qiskit 2.2.3; Qiskit Aer 0.17.2; scikit-learn 1.7.2; NumPy 2.3.5; SciPy 1.16.3; pandas 2.3.3; matplotlib 3.10.7; seaborn 0.13.2; PyTorch 2.9.1.

\section{Results}
\label{sec:results}
Unless noted otherwise, all reported results are obtained under shot-based Aer noise simulation with the same noise configuration applied to both training and evaluation.
\subsection{Aggregate continual-learning performance on UNSW-NB15 and CICIDS2017}
We report results on UNSW-NB15 and CICIDS2017 under a three-stage attack-stream with tasks
$\mathcal{T}_0$ (NORMAL vs.\ RECON\_SCAN),
$\mathcal{T}_1$ (NORMAL vs.\ DOS\_RESOURCE),
and $\mathcal{T}_2$ (NORMAL vs.\ INTRUSION\_MALWARE).
All methods share the same VQC backbone and training budget (20 epochs per task); ablations isolate the effect of (i) generative replay and (ii) stability regularization.

Tables~\ref{tab:unsw_cl_metrics} and~\ref{tab:cicids_cl_metrics} show a consistent pattern across datasets: stability regularization is the primary driver of retention, while replay is most useful as a \emph{controlled complement} once stability is enforced. 

On \textbf{UNSW-NB15}, the baseline suffers substantial forgetting (0.138). Q-FISH and EWC both move the model into a high-retention regime, with Q-FISH slightly stronger than EWC under matched budgets (mean Attack-F1 0.937 vs.\ 0.929; forgetting 0.008 vs.\ 0.013). The best overall trade-off is achieved by the full QCL-IDS configuration with gradient-space anchors (mean Attack-F1 0.941; forgetting 0.005).

On \textbf{CICIDS2017}, replay alone is more beneficial than on UNSW-NB15, improving mean Attack-F1 from 0.803 to 0.862 and reducing forgetting from 0.128 to 0.076. However, stability-based methods remain dominant: Q-FISH and EWC reach 0.927--0.936 mean Attack-F1 with markedly lower forgetting (0.021 and 0.010). The full QCL-IDS gradient-anchor variant again yields the best overall outcome (mean Attack-F1 0.944; forgetting 0.004).
\paragraph{Metrics}
We evaluate each method using both \emph{per-task classification quality} and \emph{continual-learning (CL) retention/transfer} summaries \cite{saito2015precision,davis2006relationship,he2009learning,lopezpaz2017gem}.
Because NORMAL dominates many IDS datasets, we treat the attack class ($y{=}1$) as the positive class and emphasize \textbf{Attack-F1} over raw accuracy.

Given test-set counts $\mathrm{TP},\mathrm{FP},\mathrm{FN}$ for the attack class, we compute:
\begin{equation}
\label{eq:metrics_attack_pr_re_f1}
\begin{aligned}
\mathrm{Precision}_{\mathrm{atk}}&=\frac{\mathrm{TP}}{\mathrm{TP}+\mathrm{FP}},\\
\mathrm{Recall}_{\mathrm{atk}}&=\frac{\mathrm{TP}}{\mathrm{TP}+\mathrm{FN}},\\
\mathrm{Attack\text{-}F1}&=\frac{2\,\mathrm{Precision}_{\mathrm{atk}}\,\mathrm{Recall}_{\mathrm{atk}}}{\mathrm{Precision}_{\mathrm{atk}}+\mathrm{Recall}_{\mathrm{atk}}}.
\end{aligned}
\end{equation}

We report \emph{mean Attack-F1} as the average of Attack-F1 across all tasks under the \emph{final model} after completing the full task stream.

For CL summaries, we maintain a task-performance matrix $R[t][k]$ (performance on task $k$ after training through task $t$).
Let $T$ be the number of tasks and let the diagonal entry $R[k][k]$ denote performance immediately after learning task $k$.

We report:
(i) \textbf{mean forgetting}, measuring how much earlier tasks degrade by the end of training,
\begin{equation}
\label{eq:metrics_forgetting}
\mathrm{Forgetting}=\frac{1}{T-1}\sum_{k=0}^{T-2}\Big(\max_{t\in\{k,\ldots,T-1\}} R[t][k]-R[T-1][k]\Big),
\end{equation}

(ii) \textbf{backward transfer (BWT)}, summarizing the signed change on past tasks after learning later tasks,
\begin{equation}
\label{eq:metrics_bwt}
\mathrm{BWT}=\frac{1}{T-1}\sum_{k=0}^{T-2}\big(R[T-1][k]-R[k][k]\big),
\end{equation}

and (iii) \textbf{forward transfer (FWT)}, capturing how well the model performs on a future task \emph{before} it is trained, relative to an untrained initial model (or other fixed baseline) $b_k$:
\begin{equation}
\label{eq:metrics_fwt}
\mathrm{FWT}=\frac{1}{T-1}\sum_{k=1}^{T-1}\big(R[k-1][k]-b_k\big).
\end{equation}

Unless otherwise stated, we use Attack-F1 to instantiate $R[t][k]$ and set $b_k$ to the performance of the randomly initialized model on task $k$.

\begin{table}[!ht]
\centering
\caption{UNSW-NB15 Continual learning performance.}
\label{tab:unsw_cl_metrics}
\scriptsize
\setlength{\tabcolsep}{5pt}
\renewcommand{\arraystretch}{1.0}
\begin{tabular}{lcccc}
\toprule
Method & mean Attack-F1 & Forgetting & BWT & FWT \\
\midrule
Baseline & 0.800 & 0.138 & -0.138 & 0.535 \\
QGR only & 0.791 & 0.149 & -0.148 & 0.800 \\
Conditional GMM replay & 0.806 & 0.136 & -0.134 & 0.811 \\
EWC only & 0.929 & 0.013 & -0.013 & 0.823 \\
Q-FISH only & 0.937 & 0.008 & -0.008 & 0.846 \\
QCL-IDS (state-space anchors) & 0.934 & 0.010 & -0.010 & \textbf{0.852} \\
QCL-IDS (gradient-space anchors) & \textbf{0.941} & \textbf{0.005} & \textbf{-0.005} & 0.848 \\
\bottomrule
\end{tabular}
\end{table}

\begin{table}[!ht]
\centering
\caption{CICIDS2017 continual learning performance.}
\scriptsize
\setlength{\tabcolsep}{5pt}
\renewcommand{\arraystretch}{1.0}
\label{tab:cicids_cl_metrics}
\begin{tabular}{lcccc}
\toprule
Method & mean Attack-F1 & Forgetting & BWT & FWT \\
\midrule
Baseline  & 0.803 & 0.128 & -0.128 & 0.589\\
QGR only  & 0.862 & 0.0761 & -0.076 & 0.731 \\
Conditional GMM replay & 0.841 & 0.086 & -0.086 & 0.658\\
Q-FISH only & 0.927 & 0.021 & -0.017 & 0.759\\
EWC only & 0.936 & 0.010 & -0.010 & 0.811\\
QCL-IDS (state-space anchors) & 0.933 & 0.007 & -0.007 & 0.825\\
QCL-IDS (gradient-space anchors) & \textbf{0.944} & \textbf{0.004} & \textbf{-0.004} & \textbf{0.837}\\
\bottomrule
\end{tabular}
\end{table}

\begin{table}[!ht]
\centering
\caption{Task-wise Attack-F1 scores on UNSW-NB15.}
\label{tab:unsw_attack_f1}
\scriptsize
\setlength{\tabcolsep}{5.5pt}
\renewcommand{\arraystretch}{1.0}
\begin{tabular}{lccc}
\toprule
Method & $\mathcal{T}_0$ (Recon) & $\mathcal{T}_1$ (DoS) & $\mathcal{T}_2$ (Malware) \\
\midrule
Baseline & 0.669 & 0.810 & 0.920 \\
Replay only (QGR) & 0.547 & 0.881 & 0.945 \\
Conditional GMM replay (control) & 0.604 & 0.870 & 0.944 \\
EWC only & 0.924 & 0.919 & 0.944 \\
Q-FISH only & \textbf{0.943} & 0.922 & 0.944 \\
QCL-IDS (state-space anchors) & 0.938 & 0.921 & 0.944 \\
QCL-IDS (gradient-space anchors) & 0.933 & \textbf{0.947} & \textbf{0.944} \\
\bottomrule
\end{tabular}
\end{table}

\begin{table}[!ht]
\centering
\caption{Task-wise Attack-F1 scores on CICIDS2017.}
\label{tab:cicids_attack_f1}
\scriptsize
\setlength{\tabcolsep}{5.5pt}
\renewcommand{\arraystretch}{1.0}
\begin{tabular}{lccc}
\toprule
Method & $\mathcal{T}_0$ (Recon) & $\mathcal{T}_1$ (DoS) & $\mathcal{T}_2$ (Malware) \\
\midrule
Baseline & 0.601 & 0.815 & 0.907 \\
QGR only & 0.776 & 0.884 & 0.942 \\
Conditional GMM replay & 0.736 & 0.896 & 0.937 \\
EWC only & 0.925 & 0.923 & 0.945 \\
Q-FISH only & 0.937 & 0.924 & \textbf{0.947} \\
QCL-IDS (state-space anchors) & 0.936 & 0.925 & 0.939 \\
QCL-IDS (gradient-space anchors) & \textbf{0.939} & \textbf{0.947} & 0.944 \\
\bottomrule
\end{tabular}
\end{table}

\begin{figure*}[!ht]
    \centering
    \includegraphics[width=1\linewidth]{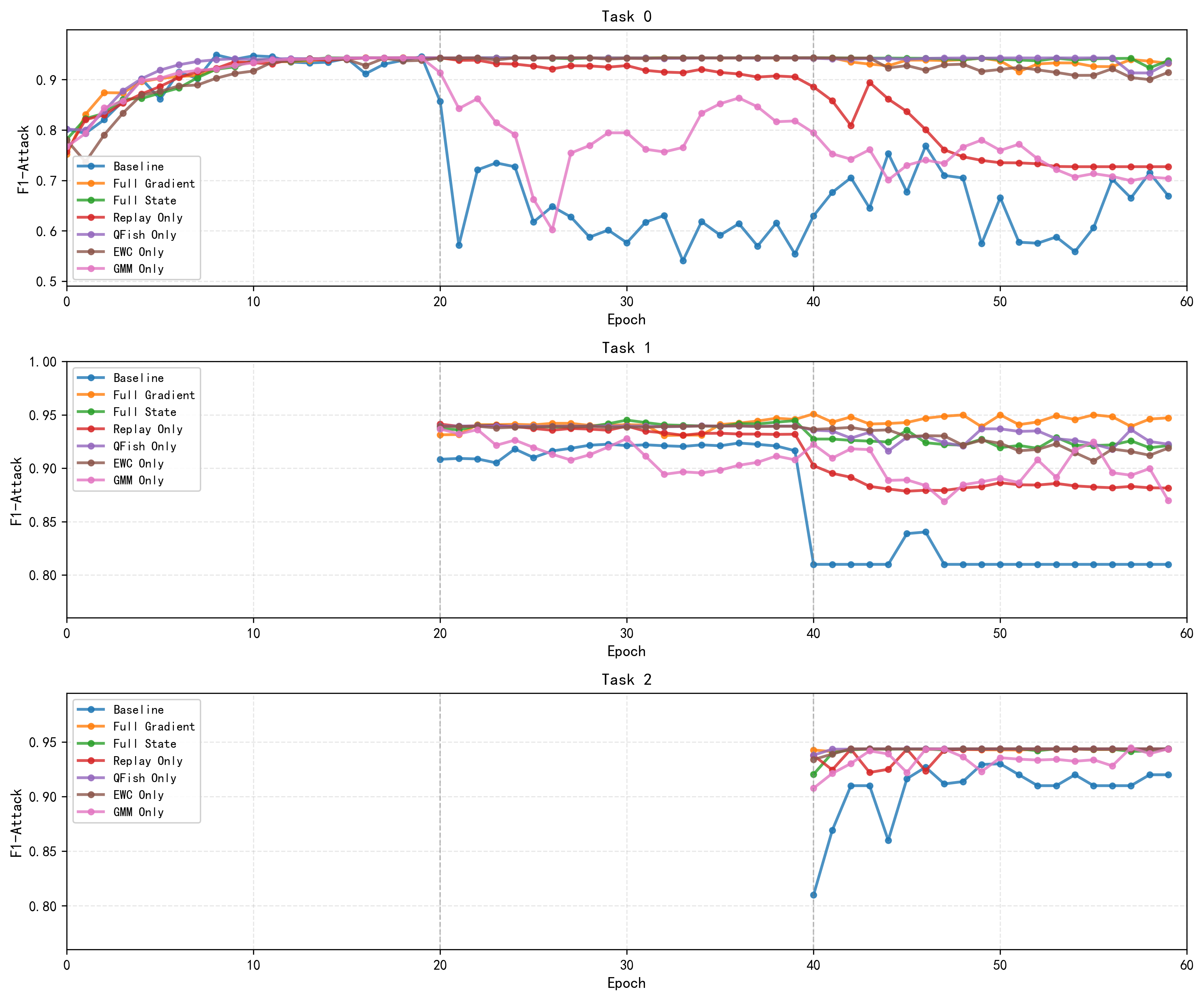}
    \caption{The curve of Attack-F1 with epoch for each task(UNSW-NB15)}
\end{figure*}

\begin{figure*}
        \centering
        \includegraphics[width=1\linewidth]{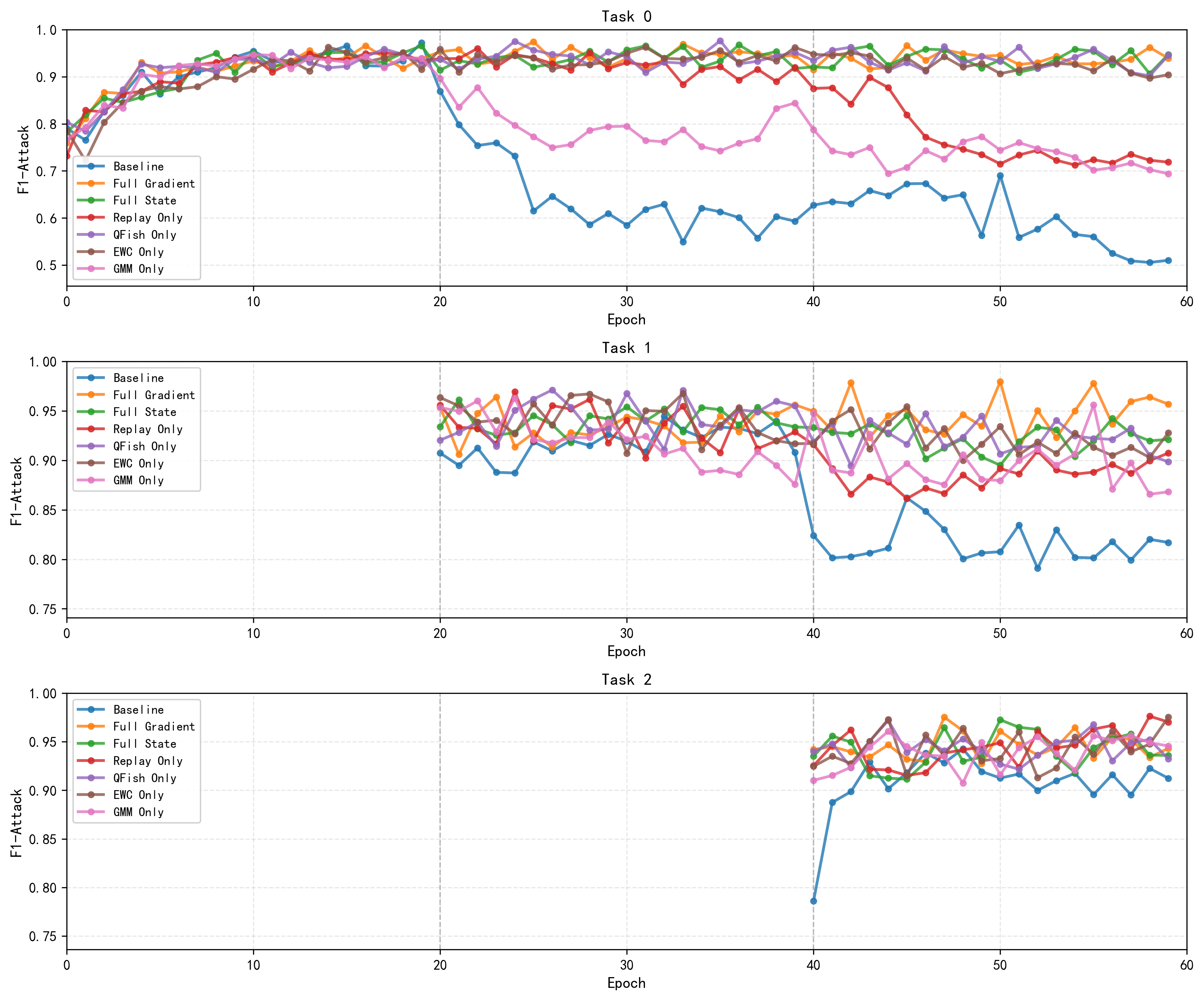}
        \caption{The curve of Attack-F1 with epoch for each task(CICIDS2017)}
    \end{figure*}

\subsection{Key finding: Q-FISH drives retention; replay alone is insufficient}
Table~\ref{tab:unsw_cl_metrics} shows that stability regularization is the dominant factor in mitigating forgetting in this stream.
The baseline exhibits substantial forgetting (0.138 mean) and poor retention on the earliest stage, with final Attack-F1 of 0.669 on $\mathcal{T}_0$.
Replay alone does not resolve this issue: \texttt{replay\_only} achieves strong late-task performance (0.945 on $\mathcal{T}_2$) but collapses on $\mathcal{T}_0$ (0.547) and yields even higher forgetting (0.149).
The classical replay control (conditional GMM) behaves similarly, with high forgetting (0.136) and weak retention on $\mathcal{T}_0$ (0.604), indicating that rehearsal without an explicit stability mechanism can induce distributional bias toward recent tasks and fail to preserve earlier-stage decision boundaries.

In contrast, Q-FISH substantially reduces forgetting while improving attack detection quality.
\texttt{qfish\_only} reaches a mean Attack-F1 of 0.937 and reduces forgetting from 0.138 to 0.008 (a 94.2\% reduction).
The full QCL-IDS variants remain in the same high-retention regime, and the gradient-anchor variant (\texttt{full\_gradient}) achieves the best overall trade-off, with mean Attack-F1 of 0.941 and the lowest forgetting (0.005), corresponding to a 96.4\% reduction relative to the baseline.

\subsection{Task-wise retention}
Task-wise results (Tables~\ref{tab:unsw_attack_f1} and~\ref{tab:cicids_attack_f1}) show that improvements are concentrated in preserving early-stage reconnaissance detection ($T_0$), which is most susceptible to catastrophic forgetting. 

For \textbf{UNSW-NB15}, replay-only collapses on $T_0$ (0.547) despite strong late-task performance on $T_2$ (0.945), indicating that rehearsal without explicit stability can bias updates toward recent tasks. In contrast, stability-regularized methods preserve $T_0$ while maintaining strong $T_2$ performance (e.g., Q-FISH-only $T_0{=}0.943$, and full variants $T_0{\approx}0.933$--0.938). 

For \textbf{CICIDS2017}, replay-only improves retention relative to baseline across all tasks ($T_0{=}0.776$ vs.\ 0.601; $T_2{=}0.942$ vs.\ 0.907), but still trails stability-regularized approaches. The full gradient-anchor model yields the strongest intermediate-stage performance on $T_1$ (0.947), while maintaining strong $T_0$ and $T_2$ performance (0.937 and 0.944). 

\subsection{Plasticity and forward transfer}
Stability regularization does not come at the expense of learning new tasks.
All Q-FISH variants exhibit strong forward transfer (FWT $\approx$0.846--0.852), substantially exceeding the baseline (0.535).
Intransigence remains small in magnitude across methods (mean $\approx$ -0.03 to -0.04), suggesting that the stabilized models are not measurably hindered in assimilating later tasks under the matched training budget.

\subsection{Efficiency and diagnostics}
Q-FISH uses a compact anchor budget (64 anchors per task), avoiding storage of historical raw data.
In exchange, it introduces moderate training overhead relative to the baseline (average training time per task increases from $\sim$2825\,s to $\sim$3523\,s for \texttt{qfish\_only}, and to $\sim$3789\,s for \texttt{full\_gradient} when combined with replay).
QFI diagnostics further differentiate anchor strategies: gradient-space anchors yield larger diagonal-QFI traces (e.g., 12.39 on $\mathcal{T}_1$ for \texttt{full\_gradient} versus 8.59 for \texttt{full\_state} and 5.19 for \texttt{qfish\_only}), consistent with sharper importance separation and the lowest observed forgetting in Table~\ref{tab:unsw_cl_metrics}.

The epoch-wise Attack-F1 curves (Figs.~5--6) qualitatively corroborate the final metrics: baseline and replay-only variants exhibit larger performance drops after task transitions, whereas Q-FISH/EWC and full QCL-IDS variants remain stable across the stream. 

Overall, the results across both datasets support the same operational conclusion: \textbf{use stability regularization as the retention backbone}, and apply replay sparingly as a secondary mechanism to improve plasticity and intermediate-stage quality once destructive drift is controlled.

\section{Discussion}
\label{sec:discussion}

This work targets a deployment-relevant continual IDS setting: sequentially assimilating emerging attack stages while retaining prior-stage detection under strict constraints on qubits (feature budget), compute, and long-term retention of raw telemetry. Our design follows the continual-learning taxonomy of combining (i) \emph{stability} mechanisms and (ii) \emph{rehearsal} mechanisms, which are widely recognized as complementary levers for mitigating catastrophic forgetting \cite{Parisi2019CLReview, vanDeVen2022ThreeTypes}. The contribution here is to instantiate these levers in a \emph{quantum-native} and \emph{storage-bounded} form suitable for NISQ-era pipelines.

\subsection{What matters most under budgeted continual IDS: stability dominates, replay complements}
Across both UNSW-NB15 and CICIDS streams, the dominant driver of retention is explicit stability regularization rather than replay alone. This mirrors a core lesson from classical CL: rehearsal without sufficient constraints can bias learning toward recent data and fail to preserve earlier decision boundaries \cite{shin2017dgr}. Concretely, replay-only improves later-stage performance but does not prevent early-stage collapse, whereas stability-regularized variants operate in a high-retention regime (low forgetting with substantially higher mean Attack-F1). These trends are consistent across datasets and persist under the same training budget, indicating that the gains are not an artifact of extra optimization steps but of improved \emph{inter-task interference control}.

Replay still matters—but primarily as a \emph{secondary} component. When paired with stability, the full systems (state- and gradient-anchor variants) yield the best overall trade-offs, with the gradient-anchor variant consistently strongest. This supports the interpretation that bounded synthetic rehearsal is most effective when it is used to \emph{support} a stable parameter trajectory, not replace it.

\subsection{Why Q-FISH is more than a “quantum EWC re-skin”}
EWC is a canonical regularization baseline that protects parameters deemed important to prior tasks \cite{kirkpatrick2017ewc}. Q-FISH inherits the high-level intent (importance-weighted constraints) but makes two substantive shifts that are especially relevant in quantum pipelines.

\textbf{(i) Functional anchoring via fidelity.} Instead of relying solely on parameter-space proximity, Q-FISH constrains \emph{behavioral drift} on a compact set of anchors through a fidelity-based term. This is closer to deployment reality: what matters operationally is preserving prior \emph{decisions} on representative traffic, not strictly preserving weights. Moreover, fidelity is a natural compatibility measure for quantum states and aligns with how variational quantum models represent data.

\textbf{(ii) Quantum-information geometry as the organizing principle.} The diagonal-QFI component can be interpreted as using a local information-geometric metric to gate updates, connecting Q-FISH to quantum natural-gradient perspectives where the quantum geometric tensor (and related Fisher quantities) defines meaningful directions in parameter space \cite{stokes2020qng}. This provides a clearer “why quantum” justification than a superficial port: the regularizer leverages structure that is intrinsic to parametrized quantum circuits (PQCs) \cite{benedetti2019parameterized}.

Importantly, classical critiques of EWC’s approximations (e.g., diagonal/online accumulation issues) motivate the need for complementary constraints \cite{Huszar2017EWCQuad}. Q-FISH’s anchor-fidelity term acts as exactly such a complement: it reduces reliance on Fisher approximations alone by adding a direct functional constraint.

\subsection{Replay under privacy and storage constraints: aligning with (but not duplicating) prior replay literature}
Generative replay is a well-established CL strategy when raw data cannot be stored. Our QGR instantiation follows this principle but adapts it to security constraints: it uses \emph{frozen, task-conditioned generator snapshots} to provide bounded rehearsal without retaining historical raw flows. Using quantum generative models (e.g., circuit Born machines) as a replay mechanism is consistent with the growing body of work treating PQCs as practical generative models on near-term devices \cite{LiuWang2018QCBM, Benedetti2018GenBenchmark}. The key distinction in our setting is not simply “quantum replay,” but \emph{replay within a strict memory-and-governance envelope}, where the generator is intentionally lightweight and used sparingly.

That said, the experiments also clarify a practical point for practitioners: replay is not a substitute for stability. In budgeted continual IDS, replay is best viewed as a controlled regularizer that improves coverage and smooths transitions \emph{after} forgetting is structurally constrained.

\subsection{Limitations and next steps}
Three limitations should be emphasized to preempt reviewer concerns. First, QFI is approximated (diagonal and anchor-conditioned), and its quality can depend on shot noise and noise models; stronger estimators or geometry-aware optimizers may further improve stability \cite{stokes2020qng}. Second, anchor selection is currently heuristic; more principled coreset construction (e.g., gradient diversity / submodular selection) could reduce anchor budget without losing coverage. Third, while generator snapshots reduce raw-data retention, synthetic replay can still inherit distributional blind spots; future work should quantify replay quality and robustness under stricter privacy threat models.

Overall, the evidence supports a concise operational takeaway: under tight feature budgets and continual attack evolution, \emph{quantum-aware stability constraints} (Q-FISH) are the primary lever for retention, while \emph{bounded} generative replay is a complementary mechanism that improves the final trade-off when coupled with stability—rather than a standalone solution.

\section{Conclusion}
\label{sec:conclusion}
This work casts continual intrusion detection as a \emph{budgeted capability-expansion} problem: IDS models must be updated as new attack stages appear, yet retain prior-stage detection under limited compute/memory and restricted retention of raw telemetry. We introduced \emph{QCL-IDS}, integrating a compact VQC classifier with (i) \emph{Q-FISH}, an anchor-based stability regularizer that couples sensitivity-weighted constraints with a fidelity-based functional drift penalty, and (ii) privacy-preserved QGR using frozen, task-conditioned generator snapshots for bounded rehearsal.

Experiments on both \textbf{UNSW-NB15} and \textbf{CICIDS2017} show a consistent operational lesson: \textbf{stability is the retention backbone, while replay is most effective as a controlled complement once drift is constrained}. In particular, the full gradient-anchor variant delivers the strongest overall trade-off (UNSW: \textbf{0.941} mean Attack-F1, \textbf{0.005} forgetting; CICIDS: \textbf{0.944} mean Attack-F1, \textbf{0.004} forgetting), while replay-only improves forward transfer but cannot prevent early-stage collapse without explicit stability control.

Future work will extend evaluation to longer, more heterogeneous streams, report multi-seed variability, and further validate robustness under noise and hardware execution, while tightening privacy guarantees for both anchor storage and replay generation.

\bibliographystyle{IEEEtran}
\bibliography{refs}

\appendix
\section{Supplementary Mathematical Details}
\label{app:math_details}

This appendix collects the detailed expressions omitted from the main text to improve readability.

\subsection{Q-FISH behavioral consistency term}
\label{app:qfish_fidelity}

A simple implementation penalizes output drift on anchors using a discrepancy between stored outputs under $\theta^{*}$ and current outputs under $\theta$:
\begin{equation}
\label{eq:qfish_fid_app}
\mathcal{L}_{\mathrm{beh}}(\theta)=
\frac{1}{|\mathcal{A}_{<t}|}\sum_{x\in\mathcal{A}_{<t}}
d\!\left(f_{\theta}(x),\,f_{\theta^{*}}(x)\right),
\end{equation}
where $d(\cdot,\cdot)$ can be mean-squared error on logits/probabilities or a divergence such as KL. The full Q-FISH regularizer is then implemented as a weighted combination of Eq.~\eqref{eq:qfi_ewc_core} and Eq.~\eqref{eq:qfish_fid_app}.

\subsection{Prototype-matching objective for generative replay}
\label{app:qgr_loss}

For a condition $c$ with prototype $\mu_c$, one convenient objective aligns measured Pauli expectations with the prototype:
\begin{equation}
\label{eq:qgr_proto_app}
\mathcal{L}_{\text{proto}}(\psi_c)=
\left\|
\mathbb{E}_{\psi_c}[Z]-\mu_{c}
\right\|_2^2.
\end{equation}
When multiple prototypes are used, the generator is trained per component and sampled as a mixture.

\subsection{Example input conditioning and re-uploading block}
\label{app:vqc_blocks}

A typical conditioning layer uses a trainable projection followed by clipping to enforce a valid encoding range:
\begin{equation}
\label{eq:linear_mixing_app}
x'=\mathrm{clip}(W x,\ a,\ b).
\end{equation}
The conditioned features are mapped to angles via a monotone transform $S(\cdot)$, and injected into each re-uploading layer along with trainable offsets, followed by an entangling block:
\begin{equation}
\label{eq:reupload_app}
R_y(\theta_{\ell,i}+S(x'_i))\ \rightarrow\ R_z(\phi_{\ell,i}).
\end{equation}
These forms are provided as one representative instantiation; the continual-learning framework is agnostic to the specific VQC ansatz used.

\end{document}